\documentclass[aps,prl,reprint,superscriptaddress]{revtex4-1}

\usepackage{graphicx}% Include figure files
\usepackage{siunitx} %SI Units
\usepackage{hyperref}
\usepackage{braket}
\usepackage{amsmath}
\usepackage{bbm}
\hypersetup{
    colorlinks=true,
    linkcolor=black,
    citecolor=black,
    filecolor=black,
    urlcolor=black,
}

\begin{document}

\title{Certifying the presence of a photonic qubit by splitting it in two}
\author{Evan Meyer-Scott}
\email{emeyersc@uwaterloo.ca}
\affiliation{Institute for Quantum Computing and Department of Physics and Astronomy, University of Waterloo, 200 University Ave W, Waterloo, Ontario, Canada N2L 3G1}
\author{Daniel McCloskey}
\affiliation{Institute for Quantum Computing and Department of Physics and Astronomy, University of Waterloo, 200 University Ave W, Waterloo, Ontario, Canada N2L 3G1}
\author{Klaudia Go\l os}
\affiliation{Institute for Quantum Computing and Department of Physics and Astronomy, University of Waterloo, 200 University Ave W, Waterloo, Ontario, Canada N2L 3G1}
\author{Jeff Z. Salvail}
\affiliation{Institute for Quantum Computing and Department of Physics and Astronomy, University of Waterloo, 200 University Ave W, Waterloo, Ontario, Canada N2L 3G1}
\author{Kent A. G. Fisher}
\affiliation{Institute for Quantum Computing and Department of Physics and Astronomy, University of Waterloo, 200 University Ave W, Waterloo, Ontario, Canada N2L 3G1}
\author{Deny R. Hamel}
\affiliation{D\'epartement de physique et d'astronomie, Universit\'e de Moncton, 18 avenue Antonine-Maillet, Moncton, New Brunswick, Canada E1A 3E9}
\author{Ad\'an Cabello}
\affiliation{Departamento de F\'{\i}sica Aplicada II, Universidad de Sevilla, E-41012 Sevilla, Spain}
\author{Kevin J. Resch}
\affiliation{Institute for Quantum Computing and Department of Physics and Astronomy, University of Waterloo, 200 University Ave W, Waterloo, Ontario, Canada N2L 3G1}
\author{Thomas Jennewein}
\affiliation{Institute for Quantum Computing and Department of Physics and Astronomy, University of Waterloo, 200 University Ave W, Waterloo, Ontario, Canada N2L 3G1}
\affiliation{Quantum Information Science Program, Canadian Institute for Advanced
Research, Toronto, ON, Canada}

\begin{abstract} 
We present an implementation of photonic qubit precertification that performs the delicate task of detecting the presence of a flying photon without destroying its qubit state, allowing loss-sensitive quantum cryptography and tests of nonlocality even over long distance. By splitting an incoming single photon in two via parametric down-conversion, we herald the photon's arrival from an independent photon source while preserving its quantum information with up to \SI{92.3\pm0.6}{\percent} fidelity. With reduced detector dark counts, precertification will be immediately useful in quantum communication.
\end{abstract}

\maketitle

{\em Introduction.}---Learning if and when a photon has arrived at a receiver without destroying it is fundamentally difficult~\cite{Grangier1998Quantum-}. 
Yet over reasonable transmission distances, this is required by advanced quantum communication protocols such as device-independent secure communication~\cite{BHK05,PhysRevLett.113.140501}, private secure randomness amplification~\cite{Colbeck06}, and fundamental tests of nonlocality~\cite{Bell64}. These proposals fail in the presence of even moderate channel losses, but can be made practical with techniques that circumvent the effects of loss by certifying that the quantum system is ready after transmission. This certification is an attractive alternative to the nearly impossible task of eliminating fundamental transmission loss mechanisms.

One approach using traditional destructive photodetection is entanglement swapping~\cite{ZZHE93}, which certifies that stationary quantum systems such as spins are ready in the desired quantum states~\cite{RWVHKCZW09}. This method has recently been applied to nitrogen-vacancy centers for a loophole-free violation of Bell's inequality~\cite{Hensen15}; however, the synchronization and indistinguishability of photons from disparate sources required for entanglement swapping is difficult to achieve, and low photon collection efficiencies from the stored quantum systems lead to low experimental rates. 

Thus we look to nondestructive optical methods of certifying a photon's arrival to indicate the photon itself is ready. One technique is heralded qubit amplification~\cite{PhysRevLett.105.070501,PhysRevA.84.022325,PhysRevA.84.010304,PhysRevA.88.012327}, which uses ancilla photons interfering with the signal photon and specific detection patterns to herald the signal's arrival. This technique requires synchronization between distant and indistinguishable photon sources and as such implementations to date have not employed a separated source and receiver~\cite{1367-2630-15-9-093002,Kocsis2013Heralded,2015arXiv150703210B}. Similarly, quantum nondemolition measurements based on cavity quantum electrodynamics and cross-phase modulation have stringent requirements on incoming photons and low repetition rates~\cite{Reiserer:2013fk,PhysRevLett.112.093601,Feizpour:2015aa,2015arXiv151001164S}.
\begin{figure}[htp]
\includegraphics[width = \columnwidth]{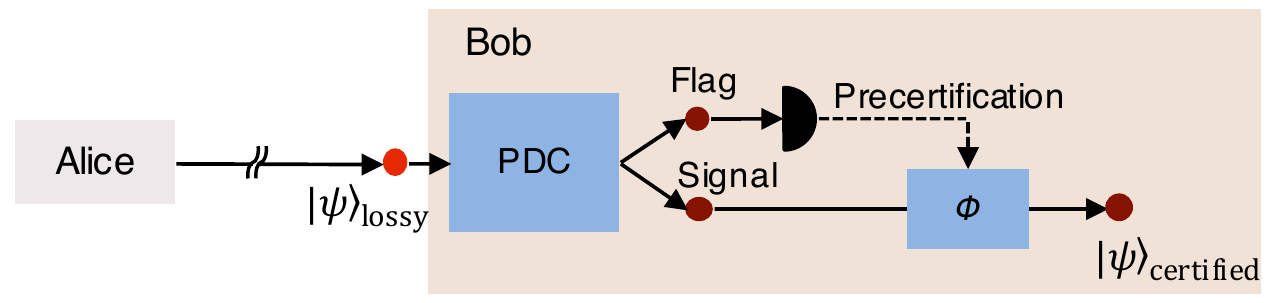} % ArXiv
\caption{(color online). Photonic qubit precertification. Alice sends a photonic qubit through a lossy channel to Bob, where it undergoes polarization-preserving parametric down-conversion (PDC), producing a photon pair with some probability. The flag photon is detected to precertify the signal, which carries the initial qubit state. For half the flag detections, a feedforward phase correction ($\phi$) is needed on the signal. \label{fig.scheme}}
\end{figure}

A promising solution for heralding the arrival of a lossy photon is photonic qubit precertification, proposed by Cabello and Sciarrino~\cite{PhysRevX.2.021010}. Precertification does not require synchronization or indistinguishability, and acts directly on flying photonic qubits. Precertification is applicable to both photonic and spin-photon entanglement, where it can overcome low coupling efficiencies out of quantum emitters.

The sender, Alice, transmits a photonic qubit encoded in polarization to the receiver, Bob. Bob splits the incoming photon into two photons, labeled flag and signal, through polarization-preserving single-photon parametric down-conversion (PDC)~\cite{Hubel2010Direct-g,Ding:15} as shown in Fig.~\ref{fig.scheme}. Since the flag photon is never produced without the corresponding signal photon, the detection of the flag precertifies the presence of the signal. The configuration is such that the signal photon bears the same quantum information initially encoded by Alice. The wavelength of Alice's photon must match the acceptance band of Bob's down-conversion crystal.

Here we report a proof-of-principle implementation of photonic qubit precertification with separated source and receiver. We demonstrate that qubit states are preserved during precertification by performing process tomography, and measurements of heralding efficiency show how our device could become useful in device-independent quantum communication.

{\em Experiment.}---Polarization-preserving PDC has the Hamiltonian
\begin{align}
H_{\text{PDC}} = \gamma \left( a_{{H}_i}^{ } a_{{H}_f}^\dagger a_{{H}_s}^\dagger + a_{{V}_i}^{ } a_{{V}_f}^\dagger a_{{V}_s}^\dagger  + H.c. \right),
\end{align}
where $\gamma$ determines the probability of down-conversion, the subscripts $i$, $f$, and $s$ refer to the input, flag, and signal modes respectively, ${H}$ and ${V}$ label two orthogonal polarization modes, and $H.c.$ means Hermitian conjugate. Unlike typical PDC the pump is a single photon and cannot be treated classically. This Hamiltonian preserves the qubit state by mapping $\ket{H}_i \rightarrow \ket{H}_f\ket{H}_s$ and $\ket{V}_i \rightarrow \ket{V}_f\ket{V}_s$. Thus an arbitrary polarization input state $\ket{\psi}_i = \alpha\ket{H}_i + \beta\ket{V}_i$ down-converts as 
\begin{align}
\alpha\ket{H}_i + \beta\ket{V} _i\rightarrow \alpha\ket{H}_f\ket{H}_s+ \beta\ket{V}_f\ket{V}_s.
\end{align}
Writing the flag mode in the diagonal basis, with $\ket{D} = \tfrac{1}{\sqrt{2}}\left(\ket{H} + \ket{V}\right)$ and $\ket{A} = \tfrac{1}{\sqrt{2}}\left(\ket{H} - \ket{V}\right)$, gives
\begin{align}
&\alpha\ket{H}_i + \beta\ket{V} _i\rightarrow \\\nonumber
&\frac{1}{\sqrt{2}}\left[ \alpha(\ket{D}_f + \ket{A}_f)\ket{H}_s+ \beta(\ket{D}_f - \ket{A}_f)\ket{V}_s\right]\\ \nonumber
&=\frac{1}{\sqrt{2}}\left[ \ket{D}_f(\alpha\ket{H}_s + \beta\ket{V}_s) + \ket{A}_f(\alpha\ket{H}_s - \beta\ket{V}_s)\right].
\end{align}
Detecting the flag qubit in $\ket{D}_f$ or $\ket{A}_f$ thus precertifies the signal qubit in the desired state, with an extra phase flip in the case of an $\ket{A}_f$ detection. The scheme works equally well if the input qubit is part of an entangled state, mapping its entanglement to the signal qubit after precertification~\cite{PhysRevX.2.021010}. This scheme does not violate the no-cloning theorem~\cite{Wootters1982A-single}, as measuring the flag photon in the diagonal basis provides no information on the coefficients $\alpha$ and $\beta$~\cite{PhysRevX.2.021010}. In fact, the input qubit state is shared across the flag and signal photons, such that neither photon in isolation can reproduce an unknown input qubit state.

\begin{figure}[htp]
\includegraphics[width = \columnwidth]{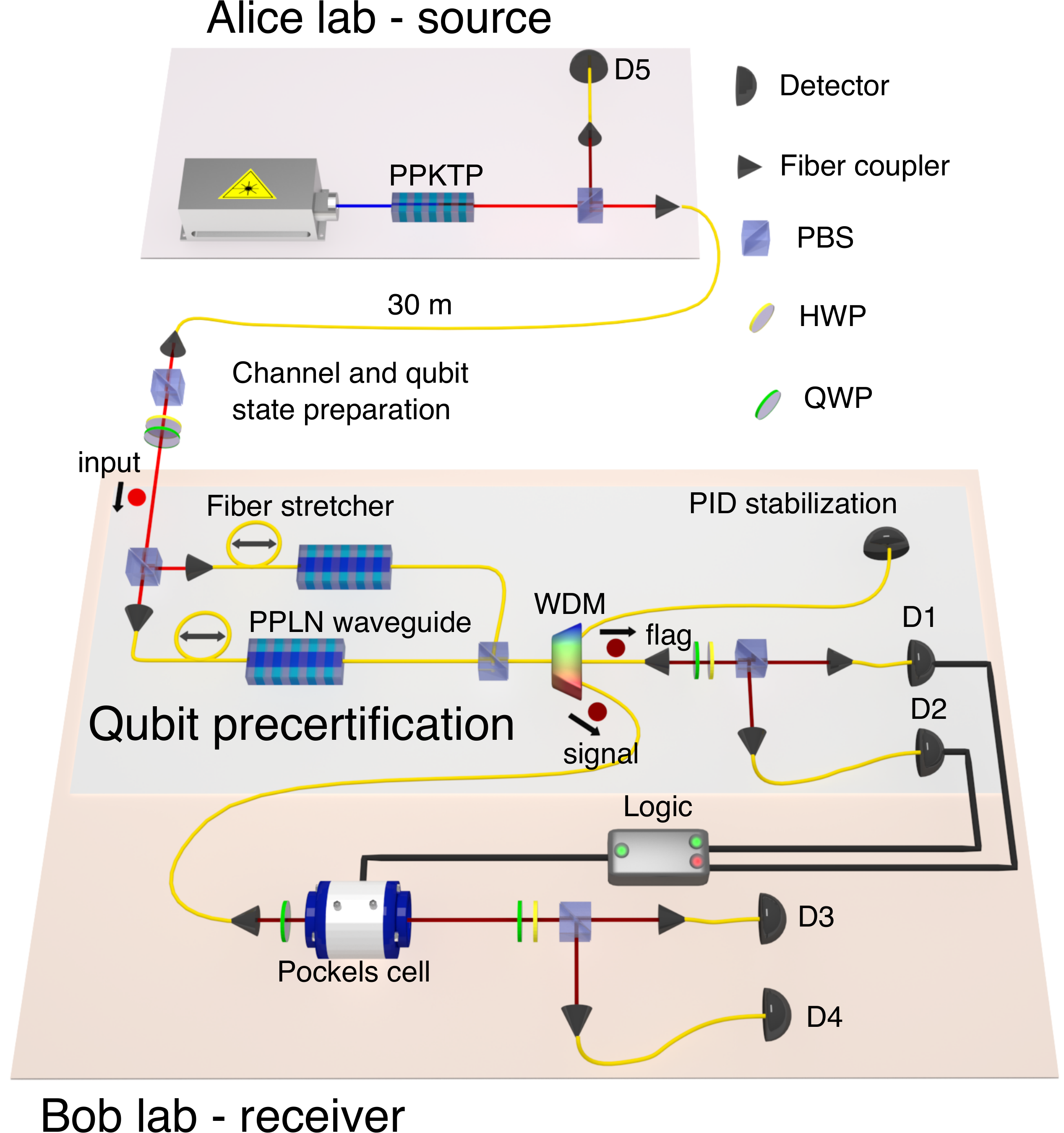} % ArXiv
\caption{(color online). Qubit precertification experiment. PPKTP - periodically-poled potassium titanyl phosphate, PBS - polarization beamsplitter, HWP - half-wave plate, QWP - quarter-wave plate, PPLN - periodically-poled lithium niobate, WDM - wavelength-division demultiplexer, PID - proportional-integral-derivative.\label{fig.setup}}
\end{figure}

Our experimental setup (Fig.~\ref{fig.setup}) comprises a source (Alice) and receiver (Bob) in separate labs with no communication except the quantum channel. Alice prepares photonic qubits through type-II down-conversion to \SI{776}{\nano\meter} wavelength in periodically-poled potassium titanyl phosphate. Alice's photons are produced in pairs with orthogonally-polarized photons, which are split off at the PBS and detected at $D5$ to enable measurement of the heralding efficiency after precertification. Alice's prepared photons are sent through a \SI{30}{\meter} optical fiber to Bob, where he performs precertification by down-converting the incoming photon in a polarization-based Mach-Zehnder interferometer, where one path down-converts $\ket{H}_i$ and the other $\ket{V}_i$ in periodically-poled type-0 lithium niobate waveguides~\cite{Tanzilli4Highly-e}. The interferometer is stabilized using a reference laser at \SI{780}{\nano\meter} to control fiber stretchers. The flag and signal photons centered at \SI{1540}{\nano\meter} and \SI{1564}{\nano\meter} respectively are split in a wavelength-division multiplexer. The flag photon is projected onto $\ket{D}_f$ or $\ket{A}_f$ with a half-wave plate, quarter-wave plate, and polarization beamsplitter, then detected with superconducting nanowire single photon detectors from Quantum Opus, LLC. The $\ket{D}_f$ or $\ket{A}_f$ outcome is fedforward to a Pockels cell, which performs the phase correction for $\ket{A}_f$ flags necessary to restore the signal photon to the input polarization state ($\ket{H}_s\rightarrow\ket{H}_s$, $\ket{V}_s\rightarrow-\ket{V}_s$). The qubit state of the signal photon is analyzed, and it is also detected by nanowire detectors. 

To record data, detections D1-D4 are registered as timetags in a coincidence logic unit from UQDevices, Inc. Alice's detection D5 (silicon avalanche photodiode) is not timetagged, but counting rates from D1-D5 are recorded on a separate logic unit to measure heralding efficiencies. The loss in count rate induced by precertification is \SI{76}{\dB}, \SI{55}{\dB} from PDC efficiency, and the rest from optical losses. We determine the loss by comparing Bob's certified pair rate ($D1\vee D2)\wedge (D3 \vee D4)$ (1100 events per hour), with the single detection rate when the qubit is measured directly by Alice (\num{1.2e7} counts per second). Here we define coincident detections using logical notation: $D_i\wedge D_j$ means logical AND between detectors $i$ and $j$, and $\vee$ means logical OR. Bob's detection efficiencies~\cite{0049-1748-10-9-A09}  are \SI{10}{\percent}, \SI{14}{\percent}, \SI{19}{\percent}, and \SI{19}{\percent}, for D1-D4 respectively, with respective dark count rates per second of \num{550}, \num{160}, \num{1500}, and \num{1000}. 

\begin{figure}[htp]
\includegraphics[width = 0.75\columnwidth]{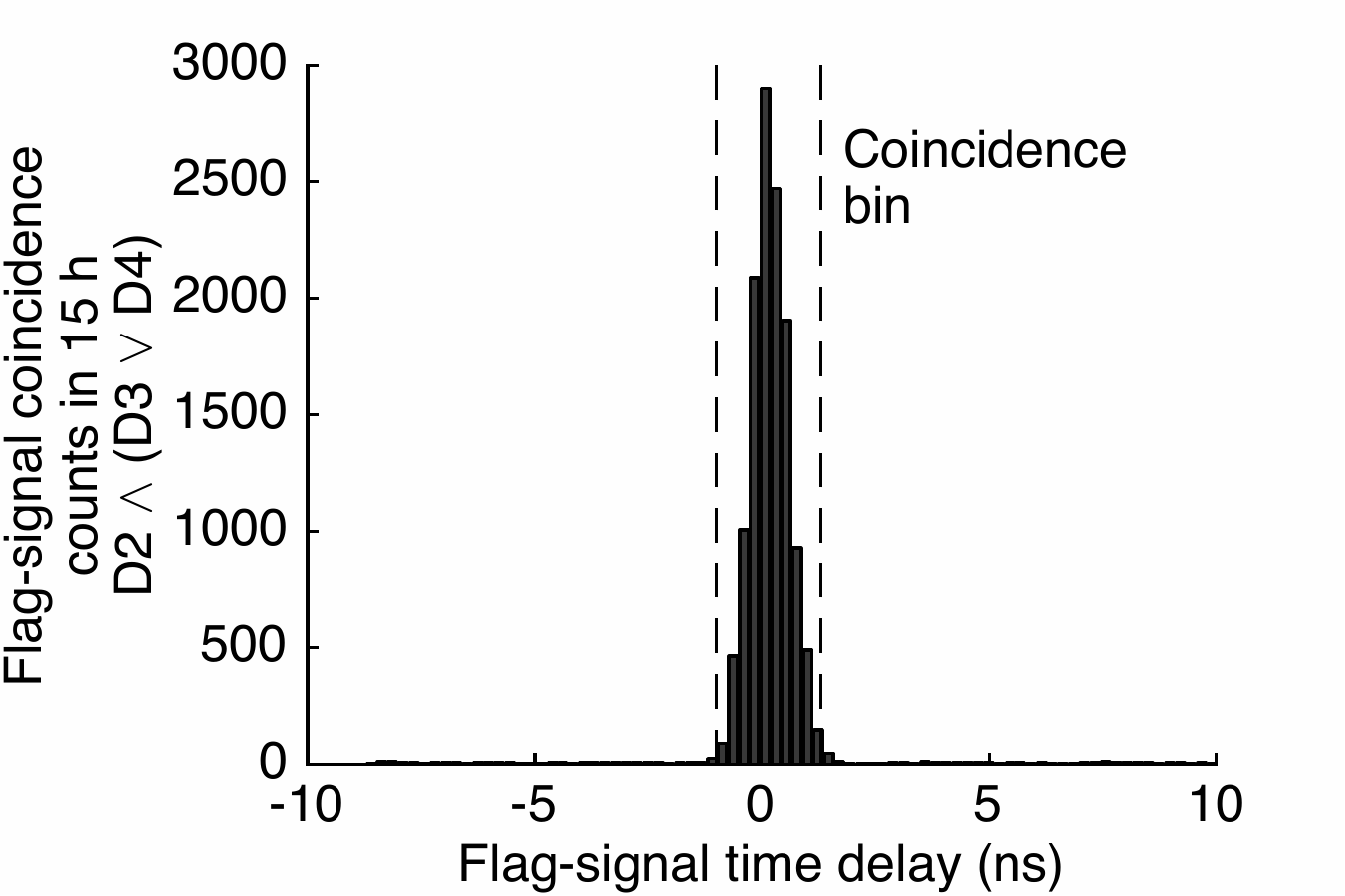} % ArXiv
\caption{Timing histogram for two-fold coincidences between Bob's flag and signal detectors $D2 \wedge (D3 \vee D4)$, for the precertification stage pumped by single photons (equivalent to \SI{1}{\pico\watt} average power). The \SI{2.3}{\nano\second} (diameter) coincidence bin used in the experiment is marked with dashed lines.\label{fig.hists}} % Run Plot_hists_pairs.m to generate this figure
\end{figure}

Interestingly the precertification stage is perhaps the weakest-pumped entangled photon pair source ever built, yet maintains a high signal-to-noise ratio (229:1) as seen in Fig.~\ref{fig.hists}.

\begin{figure}[tp]
\includegraphics[width = \columnwidth]{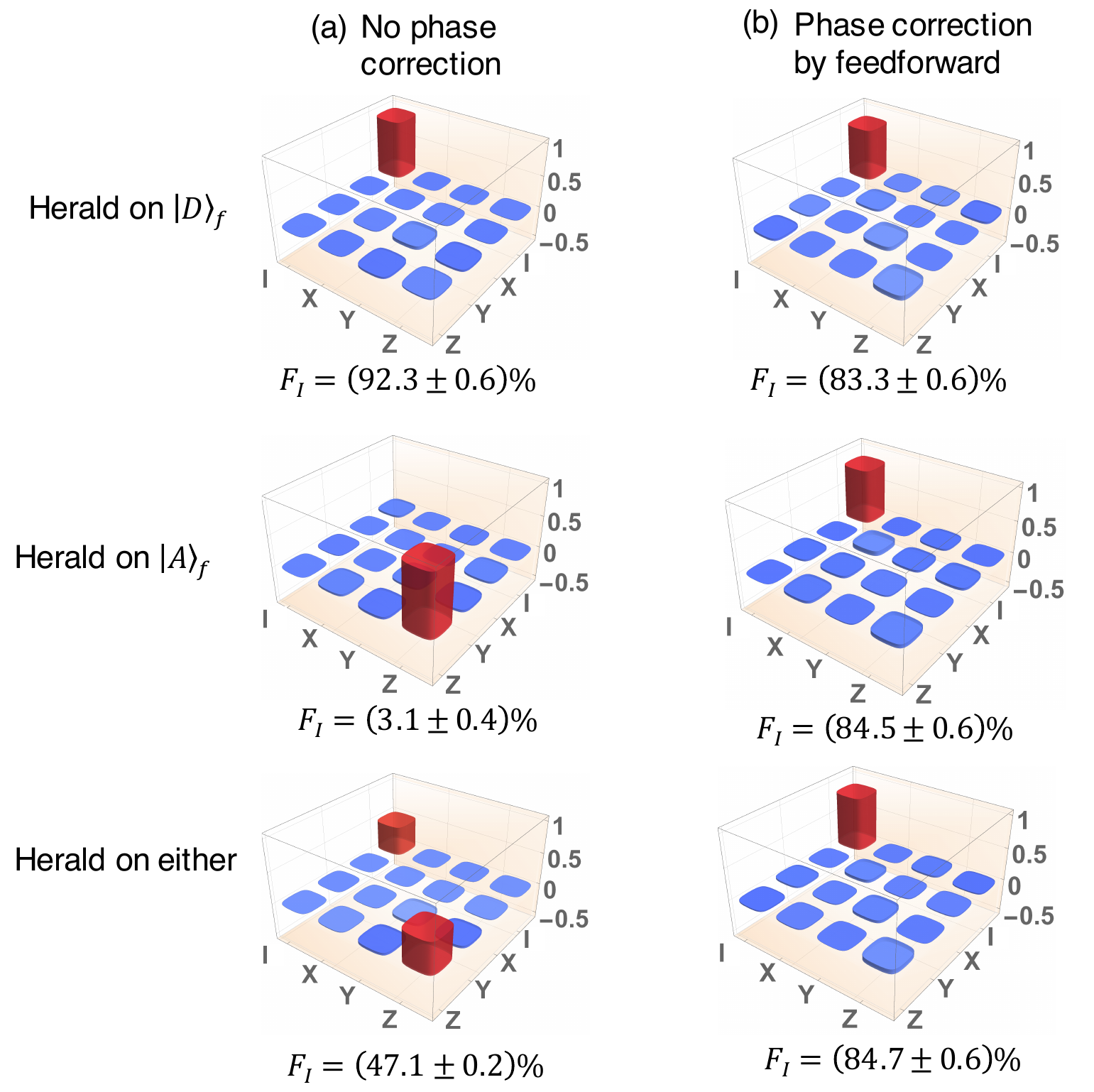} % ArXiv
\caption{(color online). Reconstruction of real parts of qubit subspace process matrices for precertification. $I$ is the identity, and $X$, $Y$, and $Z$ are the Pauli matrices. Process fidelities to the identity $F_{I}$ are given under each matrix, and imaginary parts are small (absolute values $<0.2$). (a) Without phase correction, the process fidelity depends on the flag detector. When heralding on state $\ket{A}_f$, the process has \SI{93.2\pm1.0}{\percent} fidelity to $Z$ but \SI{3.1\pm 0.4}{\percent} fidelity to $I$. Heralding on either $\ket{D}_f$ or $\ket{A}_f$ gives a mixed channel, with \SI{95.4\pm0.3}{\percent} fidelity to $(I + Z)/2$. (b) With feedforward phase correction, the process is nearly the identity independent of the flag detection pattern. \label{fig.proctomo}}
\end{figure}

{\em Results.}---To characterize the performance of qubit precertification, we performed quantum process tomography~\cite{PhysRevLett.78.390,Chuang:1997aa} in the qubit subspace by inputting 6 polarization states, and performing tomographically overcomplete measurements on the signal qubit for each input. These data were inverted by a maximum-likelihood method~\cite{PhysRevLett.93.080502} to reconstruct the process matrix~\cite{PhysRevA.64.052312}. A perfect qubit precertification would perform only the identity, directly mapping any qubit state from input to heralded output. As seen in Fig.~\ref{fig.proctomo}(a), our qubit precertification indeed performs the identity, or a phase flip for heralding on $\ket{A}_f$, with fidelities~\cite{Jozsa:1994aa} of \SI{92.3\pm0.6}{\percent} and \SI{93.2\pm1.0}{\percent} respectively. For this data set, detector $D1$ was turned off. We do not subtract background counts, and uncertainties are determined by Monte Carlo simulations assuming Poissonian counting noise. 

We can correct the phase flip using feedforward: a detection of $\ket{A}_f$ requires a $\pi$ phase, implemented by applying the half-wave voltage to the Pockels cell. Now in Fig.~\ref{fig.proctomo}(b), both $\ket{D}_f$ and $\ket{A}_f$ flag detections result in the identity on the qubit state, with \SI{84.7\pm 0.6}{\percent} fidelity. Here both detectors $D1$ and $D2$ are used. The lower fidelity is due to imperfect phase corrections of the Pockels cell.

\begin{figure}[htp]
\includegraphics[width = \columnwidth]{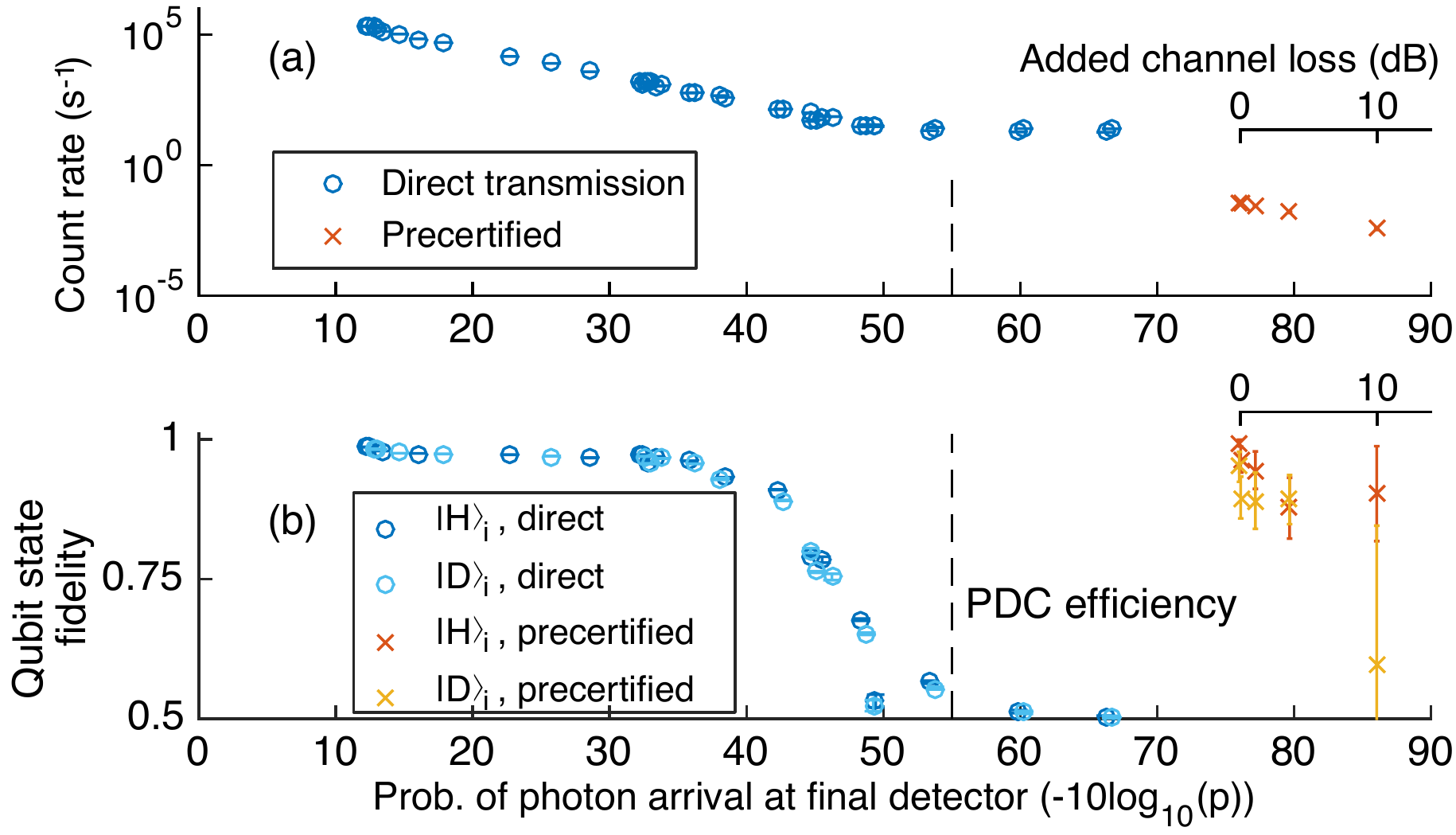} % ArXiv
\caption{(color online). (a) Count rate vs total loss, given as probability of photon arrival at final detector ($p$) in decibels, for direct transmission (o) and precertification (x). The latter includes \SI{76}{\dB} loss due to the precertification process, and the PDC efficiency from the PPLN waveguides is marked with a dashed line. The upper scales on each plot show the added channel loss for the precertification data. The direct count rate saturates around \SI{50}{\dB} loss due to dark counts.  (b) Qubit fidelities measured from quantum state tomography for input states $\ket{H}_i$  and $\ket{D}_i$. Qubits that are precertified maintain their states at much higher losses compared to those directly transmitted, which are unrecoverable after \SI{50}{\dB}. Error bars on the direct transmission data are smaller than symbol size. \label{fig.vsloss}} % Run Arrange_780_coinc_VsLoss.m to generate this figure. 
\end{figure}

We subsequently added loss to the channel between Alice and Bob to quantify how the precertified qubit state degrades and to compare with direct transmission. In Fig.~\ref{fig.vsloss} we plot count rate and Bob's measured qubit fidelity for $\ket{H}_i$ and $\ket{D}_i$ input states. Here the total loss is given in terms of the probability of a photon arriving at the final detector $p$, and includes all losses after the photon is coupled into Alice's fiber. For direct transmission, the count rate (Fig.~\ref{fig.vsloss}(a)) is the number of coincidences per second of $D5 \wedge (D1 \vee D2)$, with D1 and D2 replaced by silicon avalanche photodiodes (with the same efficiency and half the dark counts) and the precertification stage removed. With precertification, the count rate is the triple coincidences per second $D5\wedge (D1 \vee D2)\wedge (D3 \vee D4)$. As seen in Fig. \ref{fig.vsloss}(b) the qubit fidelity drops rapidly for direct transmission, whereas the precertified qubit retains a fidelity of \SI{88}{\percent} at \SI{80}{\dB} total loss or \SI{4}{\dB} added channel loss. As PDC efficiencies improve with novel materials and engineering, the precertification protocol will be able to tolerate higher channel losses while preserving qubit states.

Our rate of precertified photon-pair detection events with \SI{30}{\meter} separation and no added channel loss is \SI{0.3}{\per\second}. This is limited by the efficiency of the precertification process of \SI{-76}{\dB} (PDC efficiency of \SI{-55}{\dB}, optical coupling before PDC of \SI{-6}{\dB}, optical coupling for each the flag and signal after PDC of \SI{-5}{\dB} for a total of \SI{-10}{\dB}, and detector efficiencies for the flag of \SI{-5}{\dB}). The signal detectors have \SI{-2}{\dB} efficiency, not included in the precertification loss.

\begin{figure}[htp]
\includegraphics[width = \columnwidth]{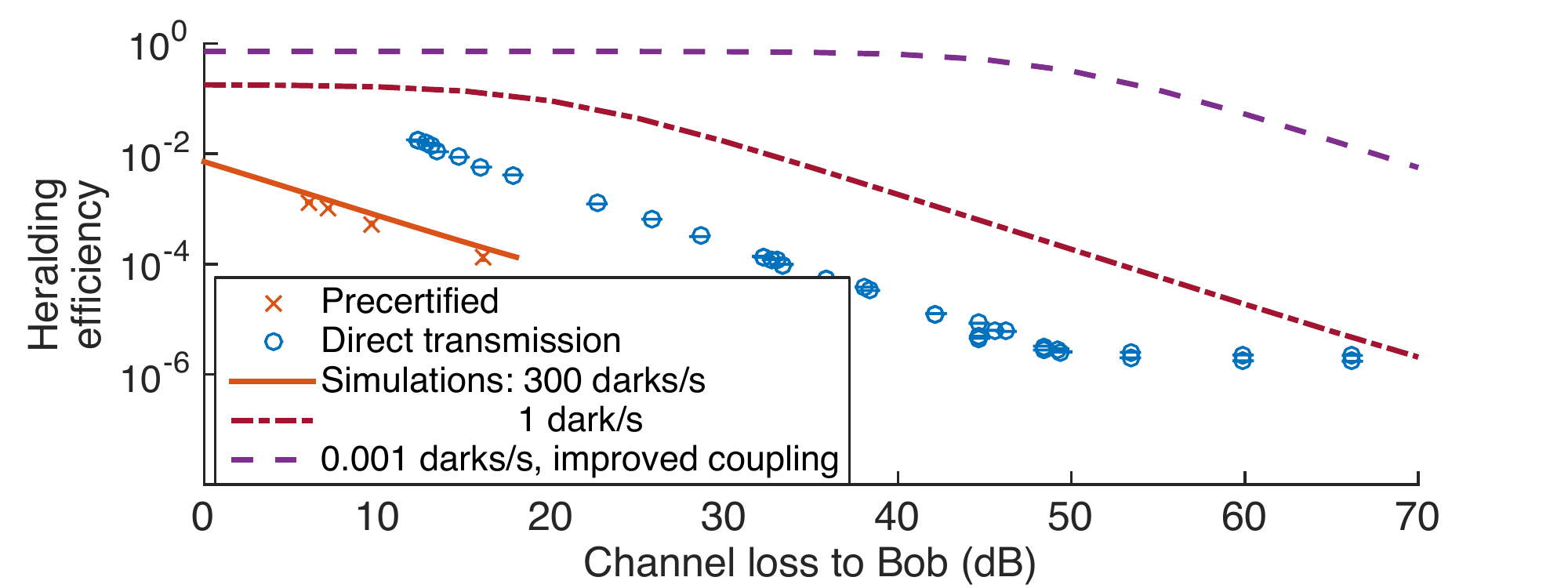} % ArXiv
\caption{(color online). Heralding efficiencies vs channel loss between Alice and Bob. The precertified heralding efficiency is limited to \num{1e-3} due to dark counts in the flag detector. The dot-dashed line shows the simulated precertification heralding efficiency given 1 dark count/s in the flag detector and \SI{100}{\pico\second} coincidence window, and the dashed line shows the additional improvement given \num{e-3} darks/s and $\eta_\mathrm{signal} =$~\SI{72}{\percent}, greatly outperforming direct transmission. \label{fig.heraldeff}} %Run Arrange_780_coinc_VsLoss.m to generate this figure. 
\end{figure}

Our measured heralding efficiencies after precertification, defined as the probability of a signal detection by Bob $(D3 \vee D4)$ given a detection by Alice $(D5)$ and a flag detection $(D1 \vee D2)$, are presented in Fig.~\ref{fig.heraldeff}. The heralding efficiency $\eta_h$ can be approximated as 
\begin{align}
\eta_h = \frac{\eta_\mathrm{signal}}{1+\frac{p_\mathrm{dark}}{p_\mathrm{flag}}},
\end{align}
where $\eta_\mathrm{signal}$ is the total efficiency of the signal photon after Bob's PDC, including coupling and detector efficiencies, and $\frac{p_\mathrm{dark}}{p_\mathrm{flag}}$ is the ratio of dark counts in the flag detector to detected flag photons, where both are triggered by Alice's $D5$ detection.
Our heralding efficiencies are limited mostly by dark counts in the flag detectors (large $\frac{p_\mathrm{dark}}{p_\mathrm{flag}}$), and partially by optical losses after precertification (small $\eta_\mathrm{signal}$). The former could be drastically improved by blackbody filters in the flag detector~\cite{Yang:14}, bringing $\eta_h \approx \eta_\mathrm{signal} = $~\SI{19}{\percent} for our system. The coupling efficiency $\eta_\mathrm{signal}$ could be improved with low-loss optical components or a chip-based architecture~\cite{Krapick:15}. Therefore in Fig.~\ref{fig.heraldeff} we also show simulated heralding efficiencies given flag detectors with \num{1}~dark count/s, \SI{10}{\percent} system efficiency, and \SI{100}{\pico\second} jitter, and also with \num{e-3} dark counts/s and \SI{2.3}{\percent} efficiency as recently demonstrated for \SI{1550}{\nano\meter}~\cite{Shibata2015Ultimate}. For the latter case we assume \SI{80}{\percent} coupling efficiency from the precertification to the detectors~\cite{PhysRevLett.93.093601,Harder:13}, and \SI{90}{\percent} efficient signal detectors~\cite{MarsiliF.:2013aa,Calkins:13}. 

For qubit precertification to be a viable alternative to entanglement swapping, the heralding efficiency must be large enough to close the detector loophole, with a lower bound of \SI{66}{\percent} for symmetric detection efficiency between Alice and Bob~\cite{Eberhard93}. The simulated heralding efficiency after precertification with \num{e-3} darks/s and improved coupling does not drop below \SI{66}{\percent} until \SI{35}{\dB} channel loss, making such a system practical for long-distance device-independent quantum communication, for example over \SI{144}{\kilo\meter} in free space~\cite{Ursin2007Entangle}. For optical fiber transmission, \SI{35}{\dB} channel loss allows \SI{10}{\kilo\meter} transmission at \SI{780}{\nano\meter}, which could be improved to \SI{175}{\kilo\meter} by moving to \SI{1550}{\nano\meter}, which could be possible using four-wave mixing instead of PDC.

{\em Conclusions.}---We have presented a proof-of-principle demonstration of photonic qubit precertification. Our device maintains the input qubit state with \SI{92.3\pm0.6}{\percent} process fidelity to the identity, or \SI{84.7\pm0.6}{\percent} with feedforward phase correction. Our heralding efficiency after precertification, up to \SI{1e-3}, can be immediately improved with low-noise detectors, and by improvements in optical components. It could also be possible to engineer phasematching in a single crystal or in a cascaded configuration~\cite{PhysRevLett.67.318} to eliminate the entanglement between flag and signal photons, and therefore remove the need for feedforward phase correction. 

Our precertification rate of \SI{0.3}{\per\second} compares favourably to state-of-the-art entanglement-swapping and quantum repeater experiments that achieved entanglement detection rates of \SI{1.3e-3}{\per\second} with \SI{3}{\meter} separation in diamond~\cite{Bernien2013Heralded-entang}, and \SI{9e-3}{\per\second} over \SI{20}{\meter} using rubidium atoms~\cite{Hofmann:2012aa}. 

 Future improvements in splitting efficiency, improving the rate and tolerable losses, are expected from advanced waveguide~\cite{PhysRevLett.108.153605} or resonator~\cite{Fortsch:2013aa} engineering, or by moving to four-wave mixing in materials that allow all operations at telecommunications wavelengths, for example in silica fibers~\cite{Li:04,PhysRevA.76.031804,PhysRevLett.99.120501}, chalcogenide fibers~\cite{Eggleton2011Chalcoge,:/content/aip/journal/apl/98/5/10.1063/1.3549744}, on-chip waveguides~\cite{Matsuda:2012aa}, or resonators~\cite{Engin:13,Grassani:15}.  These improvements and the applicability of precertification to both entangled-photon and spin-photon systems will make it a useful tool in long-distance quantum communication. 

\begin{acknowledgments}
We are grateful to Sae Woo Nam, Ant\'{\i}a Lamas Linares, and Fabio Sciarrino for helpful discussions and to Aaron Miller of Quantum Opus, LLC. We acknowledge support from the Natural Sciences and Engineering Research Council of Canada, Canada Research Chairs, Industry Canada, the Canada Foundation for Innovation, the New Brunswick Innovation Foundation, and the FQXi large grant project ``The Nature of Information in Sequential Quantum Measurements'' (A.C.), and Project No.\ FIS2014-60843-P (MINECO, Spain) with FEDER funds (AC).
\end{acknowledgments}

%\bibliography{triplets-qubit-heralder}

\end{document}